\documentclass[twocolumn,showpacs,amsmath,amssymb,superscriptaddress,aps,prc]{revtex4-1}

\usepackage{epsfig}
\usepackage{amsmath}

\usepackage{graphicx}
\usepackage{dcolumn}
\usepackage{bm}
\usepackage{mathrsfs}
\usepackage{color}

\begin{document}

\title{
Non-trivial effect for the large radius at the dripline
of Oxygen isotopes
}%

\author{N. Itagaki}

\affiliation{
Yukawa Institute for Theoretical Physics, Kyoto University,
Kitashirakawa Oiwake-Cho, Kyoto 606-8502, Japan
}

\author{A. Tohsaki}

\affiliation{
Research Center for Nuclear Physics (RCNP), Osaka University,
10-1 Mihogaoka, Ibaraki, Osaka 567-0047, Japan
}

\date{\today}

\begin{abstract}
The anomalous large radii are exotic phenomena observed around the neutron dripline.
Around the neutron dripline, 
the weak binding of the last bound neutron(s) causes the drastic increase
of the radius, which is called neutron halo structure.
Although
the nucleus $^{24}$O is located at the dripline of Oxygen isotopes,
the separation energies of one and two neutron(s) are 4.19 MeV and  6.92 MeV, respectively.
In spite of this enough binding,
the enhancement of the matter radius is observed.
In this study, we microscopically describe the structure change of $^{22}$O
core in $^{24}$O and explain the observed large radius based on the cluster model.
Two degrees of freedom for the large radius; the relative distances among
four $\alpha$ clusters and size of each $\alpha$ cluster
are examined,
where Tohsaki interaction, which has finite range three-body terms
is employed. The nucleus $^{24}$O has the almost the same amount of the clusterization 
compared with $^{22}$O, but the expansion of each $\alpha$ cluster plays an important role.
When two neutrons are added to $^{22}$O at the center,
the expansion of each $\alpha$ clusters is energetically more favored than enhancing the clustering 
for reducing the kinetic energy of the neutrons.
The calculated rms matter radius of $^{22}$O and $^{24}$O are 2.75 fm and 2.92 fm, respectively.
Although these are slightly smaller than the experimental values,
the jump at $^{24}$O from $^{22}$O is reproduced. 

\end{abstract}

\maketitle

\section{Introduction}
The anomalous large radii are exotic phenomena observed around the neutron dripline.
This is 
never seen in normal nuclei, where nuclear saturation plays an important 
and nuclear density and energy density are rather constant.
Around the neutron dripline, the ratio of protons and neutrons is not optimal,
which results in the weak binding of the last bound neutron.
In $^{11}$Li, the binding energy of the last two neutrons is only 0.36936 MeV
compared with typically 8 MeV per nucleon in normal nuclei, and these two neutrons
spread with large radius.
This is called neutron halo structure, and the matter radius of $^{11}$Li was deduced to be 
3.27$\pm$0.24 fm~\cite{PhysRevLett.55.2676}, much larger than the charge radius of 
 2.467(37) fm~\cite{PhysRevLett.96.033002}.
This large radius is attributed to the tunnel effect of 
two valence neutrons, which are loosely bound around
the $^9$Li core.

The nucleus $^{24}$O is also located at the dripline and it is one of the isotones corresponding  
to the new magic number of $N=16$ for the neutrons~\cite{PhysRevLett.84.5493}.
Although this is a dripline nucleus, the binding of neutron(s) is rather strong;
the separation energies of one and two neutron(s) are 4.19 MeV and  6.92 MeV, respectively.
The last valence neutrons are ``well bound'' compared with other light dripline nuclei.
In spite of this enough binding, it is quite surprising that 
the enhancement of the matter radius is observed for $^{23}$O and $^{24}$O~\cite{OZAWA2001599}.
The deduced matter radii of $^{23}$O and $^{24}$O are 3.20$\pm$0.04 fm and 3.19$\pm$0.13 fm, respectively,
whereas $^{22}$O has much smaller value of 2.88$\pm$0.06 fm. 
Since $^{23}$O and $^{24}$O are not considered to have the neutron halo structure because of the enough 
amount of separation energies of the neutron(s), there must exist non-trivial effect for the large radii.

Until now various mechanism was proposed for the non-trivial large radius.
One of the idea is the modification of the $^{22}$O core;
the size of $^{22}$O in $^{24}$O may get larger than the free one
owing to the additional two neutrons~\cite{PhysRevLett.88.142502,PhysRevC.73.034318}.
Indeed, the large radius of $^{24}$O can be explained by assuming the increase of the $^{22}$O core size.
Here the question is how this change of the core size happens.
The roles of the nature of realistic interactions have been discussed~\cite{PhysRevLett.108.242501,PhysRevLett.117.052501}.

It has been widely known that the structure of $^{16}$O is well described by 
four $\alpha$ models~\cite{PhysRevC.29.1046,PhysRevLett.81.5291,Bijker2017154}. 
The binding energy per nucleon of $^4$He is quite large in light mass region, thus
the $\alpha$ particles 
are considered as good building blocks of the nuclear structure called $\alpha$ clusters~\cite{PTPS.68.29}.
In $^{16}$O, the tetrahedron configuration of four $\alpha$ clusters has been known to
coincide with the doubly closed shell structure of the $p$ shell at the 
small distance limit between the $\alpha$ clusters owing to the antisymmetrization effect.
Moreover, the energy optimal state has finite distance between $\alpha$ clusters,
which is proven also by recent {\it ab initio} studies~\cite{NEFF2004357,PhysRevLett.112.102501}.
Also, it is discussed that gas-like state of four $\alpha$ clusters appears
around the corresponding threshold energy, which is analogous to the famous Hoyle state (three $\alpha$ state)
in $^{12}$C~\cite{PhysRevLett.87.192501}.

The purpose of the present study is to describe the structure change of $^{22}$O
core in $^{24}$O and explain the observed large radius of $^{24}$O based on the cluster model.
There two degrees of freedom for the large radius; the relative distances among
the $\alpha$ clusters and size of each $\alpha$ cluster
are variationally determined.

For such calculations, we need a reliable interaction, which acts among the nucleons.
It is quite well known that
the central part of the interaction 
should have proper density dependence in order
to satisfy the saturation property of nuclear systems.
If we just introduce simple
two-body interaction, for instance Volkov interaction~\cite{VOLKOV196533}, which has been
widely used in the cluster studies, we have to
properly choose Majorana exchange parameter for each nucleus, and consistent description of
two different nuclei with the same Hamiltonian becomes a tough work.
Adding zero-range three-body interaction term helps better agreements with
experiments; however
the radius and binding energy of free $^4$He ($\alpha$ cluster) are not well reproduced.
The Tohsaki interaction, which has finite range three-body terms, has much advantages~\cite{PhysRevC.49.1814,PhysRevC.94.064324}.
Although this is phenomenological interaction,
it gives reasonable size and binding energy of the $\alpha$ cluster,
and $\alpha$-$\alpha$ scattering phase shift is reproduced,
while
the saturation properties of nuclear matter is also reproduced rather satisfactory.
Thus we adopt this interaction.

One of the problems of the traditional cluster models is that
the spin-orbit  interaction,
quite important in explaining the observed magic numbers,
does not contribute inside $\alpha$ clusters and also between $\alpha$ clusters.
In cluster models, each $\alpha$ cluster is often defined as a simple $(0s)^4$ configuration 
at some spatial point, which is spin singlet free from
the non-central interactions.
To include the spin-orbit contribution
starting with the cluster model,
we proposed the antisymmetrized quasi-cluster model 
(AQCM)~\cite{PhysRevC.94.064324,PhysRevC.73.034310,PhysRevC.75.054309,PhysRevC.79.034308,PhysRevC.83.014302,PhysRevC.87.054334,ptep093D01,ptep063D01},
which allows smooth transition of $\alpha$ cluster model wave function to
$jj$-coupling shell model one.
We call the clusters which are transformed to feel the spin-orbit effect quasi clusters.
In AQCM, 
we have only two parameters: $R$ representing the distance between $\alpha$ clusters
and $\Lambda$, which characterizes the transition of $\alpha$ cluster(s) to quasi cluster(s).
It has been known that the conventional $\alpha$ cluster models cover the model space of closure of major shells
($N=2$, $N=8$, $N=20$, {\it etc.}) of the $jj$-coupling shell model.
In addition, we have shown that 
the subclosure configurations of the $jj$-coupling shell model,
$p_{3/2}$ ($N=6$), $d_{5/2}$ ($N=14$), $f_{7/2}$ ($N=28$), and $g_{9/2}$ ($N=50$)
can be described by our AQCM~\cite{ptep093D01}.

In the present case, the  $^{16}$O core part can be described within the
four $\alpha$ cluster model. The four $\alpha$ cluster wave function
with a tetrahedron configuration
coincides with the lowest configuration of the $jj$-coupling shell model 
(doubly closed configuration of the $p$ shell)
when $\alpha$ clusters get closer. 
In the case of doubly closed configuration,
both spin-orbit attractive and repulsive orbits
are filled, and the spin-orbit interaction does not contribute;
introducing $\alpha$ cluster models free from the spin-orbit interaction
does not harm.
However, the neutrons outside of the $^{16}$O core are in the $sd$ shell, and
the spin-orbit effect has to be treated for them.
Since the di-neutron component is known to be important in dripline region~\cite{PhysRevC.78.017306},
we introduce di-neutron configurations for these neutrons, which are free from
the spin-orbit effect, and then, they are transformed to quasi clusters by giving $\Lambda$ parameter,
and the contribution of the spin-orbit interaction can be taken into account.


\section{The Model\label{model}}
%

The Hamiltonian ($\hat{H}$) consists of kinetic energy ($\hat{T}$) and 
potential energy ($\hat{V}$) terms,
\begin{equation}
\hat{H} = \hat{T} +\hat{V},
\end{equation}
and the kinetic energy term is described as one-body operator,
\begin{equation}
\hat{T} = \sum_i \hat{t_i} - T_{cm},
\end{equation}
and the center of mass kinetic energy ($T_{cm}$),
which is constant,
is subtracted.
The potential energy has
central ($\hat{V}_{central}$), spin-orbit ($\hat{V}_{spin-orbit}$), 
and the Coulomb parts.

For the central part of the potential energy
($\hat{V}_{central}$), the Tohsaki interaction is adopted,
which consists of two-body ($V^{(2)}$)  and three-body ($V^{(3)}$) terms:
\begin{equation}
\hat{V}_{central} = {1 \over 2} \sum_{i \neq j} V^{(2)}_{ij} 
+ {1 \over 6} \sum_{i \neq j, j \neq k, i \neq k}  V^{(3)}_{ijk},
\end{equation}
where $V^{(2)}_{ij}$ and $V^{(3)}_{ijk}$ consist of three terms with different range parameters,
\begin{equation}
V^{(2)}_{ij} =  
\sum_{\alpha=1}^3 V^{(2)}_\alpha \exp[- (\vec r_i - \vec r_j )^2 / \mu_\alpha^2]
 (W^{(2)}_\alpha + M^{(2)}_\alpha P^r)_{ij},
\label{2body}
\end{equation} 
\begin{eqnarray}
V^{(3)}_{ijk} = 
\sum_{\alpha=1}^3 && V^{(3)}_\alpha \exp[- (\vec r_i - \vec r_j )^2 / \mu_\alpha^2 -
                                                     (\vec r_i - \vec r_k)^2 / \mu_\alpha^2   ] \nonumber \\
\times &&  (W_\alpha^{(3)} + M_\alpha^{(3)} P^r)_{ij} (W_\alpha^{(3)} + M_\alpha^{(3)} P^r)_{ik}.
\end{eqnarray}
Here, $P^r$ represents the exchange of spatial part of the wave functions
of interacting two nucleons.
In this article, we use
F1 parameter set.

For the spin-orbit part,
G3RS \cite{PTP.39.91}, which is a realistic
interaction originally determined to reproduce the nucleon-nucleon scattering phase shift, 
is adopted;
\begin{equation}
\hat{V}_{spin-orbit}= {1 \over 2} \sum_{i \ne j} V^{ls}_{ij},
\end{equation}
\begin{equation}
V^{ls}_{ij}= V_{ls}( e^{-d_{1} (\vec r_i - \vec r_j)^{2}}
                    -e^{-d_{2} (\vec r_i - \vec r_j)^{2}}) 
                     P(^{3}O){\vec{L}}\cdot{\vec{S}}.
\label{Vls}
\end{equation}
The strength of the spin-orbit interaction, $V_{ls}$,  is the only parameter 
in the present Hamiltonian, and
$V_{ls} = 2000$ MeV is adpted, which has been tested in many 
earlier studies~\cite{PhysRevC.61.044306,PhysRevC.70.054307}, 
including our former calculation for $^{18}$O~\cite{PhysRevC.79.034312}.


The wave function of the total system $\Psi$ is antisymmetrized product of
single particle wave functions;
\begin{equation}
\Psi = {\cal A} \{ \psi_1 \psi_2 \psi_3 \cdot \cdot \cdot \cdot \psi_A \},
\label{total-wf}
\end{equation}  
where $A$ is a mass number.
The single particle wave function has a Gaussian shape 
as in the conventional $\alpha$ cluster models;
\begin{align}
        \psi_{i} &= \left( \frac{2\nu}{\pi} \right)^{\frac{3}{4}}
                \exp \left[- \nu \left(\bm{r}_{i} - \bm{\zeta}_{i} \right)^{2} \right] \chi_{i} \tau_{i}.
\label{spwf} 
\end{align}
where $\chi_{i}$ and $\tau_{i}$ in Eq.~\eqref{spwf} represent the spin and isospin part of the $i$-th
single particle wave function, respectively.

\begin{figure}[t]
	\centering
	\includegraphics[width=6.5cm]{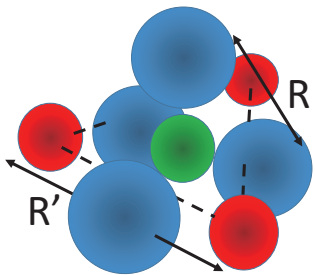} 
	\caption{
Schematic figure for $^{24}$O.
Large blue spheres show the four $\alpha$ clusters with the relative distance of $R$ (fm)
for the $^{16}$O core. The small red spheres are di-neutron clusters with the relative 
distance of $R'$ (fm), which are changed to quasi clusters based on AQCM.
One more di-neutron (small green sphere) is added at the origin corresponding to $(2s)^2$.
\label{schema}
     }
\label{ft}
\end{figure}

The schematic configuration is shown in Fig.~\ref{schema}.
For the nucleons $i=1 \sim 16$,
they correspond to the $^{16}$O core,
and we introduce a tetrahedron configuration of four $\alpha$ clusters (large blue spheres). 
The distances between the centers of $\alpha$ clusters are characterized by the parameter $R$ (fm). 
The same as in Brink-Bloch wave function, 
four nucleons in one $\alpha$ cluster share a common value for the Gaussian center parameter
$\bm{\zeta}_{i}$, 
and the contribution of the spin-orbit interactions vanishes for the $^{16}$O core.
This is also true in the $jj$-coupling shell model;
$^{16}$O corresponds to double closed,
where both $j$-upper and $j$-lower orbits are filled.
The wave function of four $\alpha$ clusters coincides with the shell model at the small limit of $R$. 

On the other hand, 
the eight valence neutrons around $^{16}$O ($i=17 \sim 24$)
must be introduced so that 
the contribution of the spin-orbit interaction can be estimated.
Six of them (small red spheres) are introduced as three di-neutron
clusters with an equilateral triangular shape. The relative distance is characterized with $R'$ (fm).
There is no spin-orbit effect for the di-neutron clusters,
thus
they are changed into quasi clusters based on AQCM.
When the original position of the cluster is at $\bm{R}$,
the Gaussian center parameter of each neutron in this cluster is transformed as
\begin{equation}
\bm{\zeta} = \bm{R} + i \Lambda \bm{e}^{\text{spin}} \times \bm{R}, 
\label{AQCM}
\end{equation}
where $\bm{e}^{\text{spin}}$ is a unit vector for the intrinsic-spin orientation of this neutron,
and $\Lambda$ is a real control parameter for the imaginary part.
Now we place one di-neutron cluster on the $x$ axis and transform it 
to quasi cluster based on AQCM.
The Gaussian center parameter
of the spin-up neutron is transformed as
\begin{equation}
\bm{\zeta}_{17} = R'(\bm{e_x} + i \Lambda \bm{e_y})/\sqrt{3},
\end{equation}
and for the spin-down neutron, it is transformed as
\begin{equation}
\bm{\zeta}_{18} = R'(\bm{e_x} - i \Lambda \bm{e_y})/\sqrt{3},
\end{equation}
where ${\bf e_x}$ and ${\bf e_y}$ are unit vectors in the $x$ and $y$ direction,
respectively.
By introducing the imaginary part, we can describe the time reversal motion
of the two neutrons, and the spin-orbit interaction contributes.
The second and third quasi clusters
are introduced by rotating both spatial and spin parts
of these spin-up and down neutrons around the $y$-axis
by 120$^o$ and 240$^o$, respectively.
These six neutrons becomes $(d_{5/2})^6$ configuration 
at $R' \to 0$ and $\Lambda  = 1$.
For $^{24}$O,
we further add two neutrons, and here
we add one di-neutron cluster at the center 
of the system (small green sphere in Fig.~\ref{schema}), 
which becomes $(2s)^2$ configuration owing to the antisymmetrization effect
at $R \to 0$.
The parameters $R'$ and $\Lambda$ are variationally determined 
for each $R$ value.

\section{Results}

\begin{figure}[t]
	\centering
	\includegraphics[width=6.5cm]{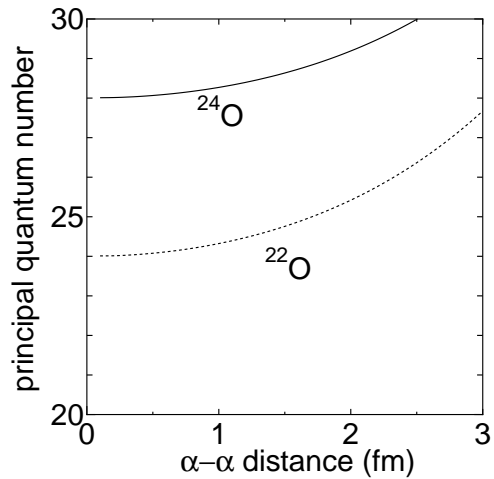} 
	\caption{
The principal quantum number of  $^{22}$O and $^{24}$O
as a function of $\alpha$-$\alpha$ distance for the four $\alpha$ clusters
($R$ in Fig..~\ref{schema}).
The size parameters $\nu$ in Eq.~\ref{spwf}  for $^{22}$O and $^{24}$O are
0.18 fm$^{-2}$ and 0.16 fm$^{-2}$, respectively.
     }
\label{o22-24-n}
\end{figure}

\begin{figure}[t]
	\centering
	\includegraphics[width=6.5cm]{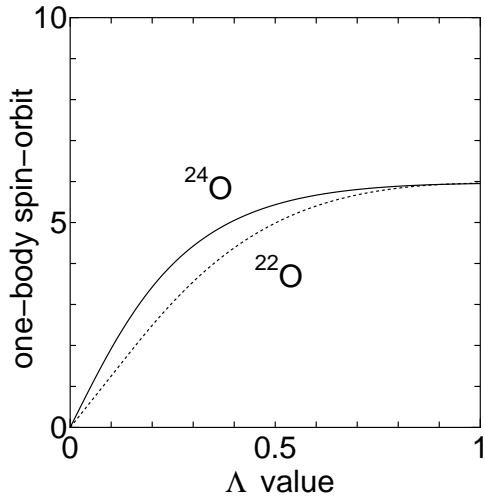} 
	\caption{
Expectation values of one-body spin-orbit operator $\sum_i {\bf l_i} \cdot {\bf s_i}$ of $^{20}$O and $^{24}$O
as a function of $\Lambda$ value at the limit of $R$, $R' \to 0$.
The size parameters $\nu$ in Eq.~\ref{spwf}  for $^{22}$O and $^{24}$O are
0.18 fm$^{-2}$ and 0.16 fm$^{-2}$, respectively.
     }
\label{o22-24-ls}
\end{figure}

As mentioned before,
the AQCM wave function includes the lowest configuration of the $jj$-coupling shell model wave function.
Three di-neutrons around $^{16}$O
are changed into the $(d_{5/2})^6$ configuration.
This can be analytically proven, but here we numerically show it.
The expectation values of
 principal quantum number
of the harmonic oscillator ($n$)
is shown in Fig.~\ref{o22-24-n}.
When 
the $\alpha$-$\alpha$ distance for the four $\alpha$ clusters
with the tetrahedron shape ($R$ in Fig.~\ref{schema}) reaches zero,
the values converge to 24 ($^{22}$O) and 28 ($^{24}$O), 
where 4 nucleons are in the lowest $s$ shell ($n=0)$, 12 nucleons are in the $p$ shell ($n=1$),
and
 6 and 8 neutrons are in the $sd$ shell ($n = 2$) for $^{22}$O and $^{24}$O, respectively.
Also, Fig.~\ref{o22-24-ls} shows
the expectation values of  the one-body spin-orbit operator $\sum_i {\bf l_i} \cdot {\bf s_i}$
in the unit of $\hbar^2$,
as a function of $\Lambda$ in Eq.~\ref{AQCM}.
This  is calculated at the shell model limit ($R , R' \to 0$),
and the value is zero at $\Lambda = 0$ but it becomes 6 at $\Lambda =1$.
This means that three di-neutron clusters are changed into
six neutrons in the $d_{5/2}$ orbits of the $jj$-coupling shell model,
since the eigen value of ${\bf l_i} \cdot {\bf s_i}$ for one neutron
in $d_{5/2}$ is $\{j(j+1)-l(l+1)-s(s+1) \}/2 = \{ 35/4 -6 - 3/4\}/2 = 1$.

\begin{figure}[t]
	\centering
	\includegraphics[width=6.5cm]{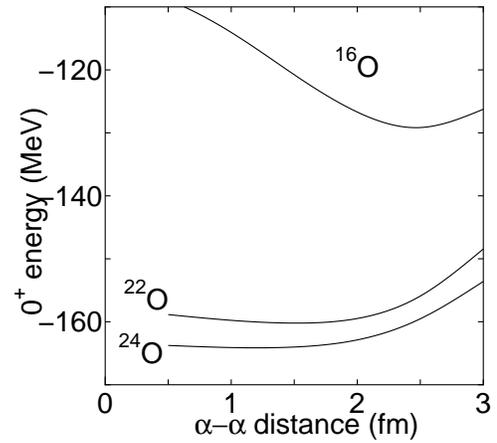} 
	\caption{
The $0^+$ energies of $^{16}$O, $^{20}$O and $^{24}$O
as a function of $\alpha$-$\alpha$ distance for the four $\alpha$ clusters
($R$ in Fig.~\ref{schema}).
The size parameter $\nu$ of the single part wave function in Eq.~\ref{spwf}
is chosen so as to obtain the lowest energy; $\nu = 0.23$ fm$^{-2}$ for $^{16}$O, 
$\nu = 0.18$ fm$^{-2}$ for $^{22}$O, 
and
$\nu = 0.16\ {\rm fm}^{-2}$ for $^{24}$O.
     }
\label{energy-curve}
\end{figure}

The projections of the wave functions onto parity and angular momentum eigenstates can be performed 
numerically, and
the $0^+$ energy curves for $^{16}$O, $^{22}$O, and $^{24}$O
are shown in Fig.~\ref{energy-curve}
as a function of $\alpha$-$\alpha$ distance for the four $\alpha$ clusters
($R$ in Fig.~\ref{schema}).
The parameters $R'$ and $\Lambda$ are optimized for each $R$, and the adopted values
for $^{24}$O is summarized in Table~\ref{adopted-value}.
Here, the size parameter $\nu$ of the single part wave function in Eq.~\ref{spwf}
is chosen so as to obtain the lowest energy; $\nu = 0.23$ fm$^{-2}$ for $^{16}$O, 
$\nu = 0.18$ fm$^{-2}$ for $^{22}$O, 
and
$\nu = 0.16\ {\rm fm}^{-2}$ for $^{24}$O.
The $^{16}$O nucleus has rather clear energy minimum point around 
$R = 2-2.5$ fm, but with increasing number of neutrons, 
the energy curves become more flat at small $R$ regions.  
Nevertheless, $^{22}$O and $^{24}$O have almost the same clustering features;
the energy minimum points appear around $R = 1.0-2.0$ fm region.

\begin{figure}[t]
	\centering
	\includegraphics[width=6.5cm]{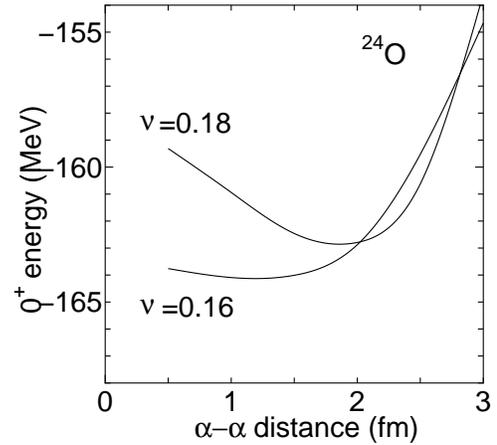} 
	\caption{
The size parameter $\nu$ dependence for the $0^+$ energy of $^{24}$O
as a function of $\alpha$-$\alpha$ distance for the four $\alpha$ clusters
($R$ in Fig.~\ref{schema}).
     }
\label{nu-dep}
\end{figure}

The $\nu$ parameter dependence of $^{24}$O is shown in Fig.~\ref{nu-dep},
and $\nu = 0.16\ {\rm fm}^{-2}$ gives lower energy compared with  $\nu = 0.18\ {\rm fm}^{-2}$,
which is the optimal value for $^{22}$O.
Using the relation of $\nu = 1/2b^2$, 
sizable effect of the expansion of each $\alpha$ cluster in $^{24}$O can be seen;
$\nu = 0.18\ {\rm fm}^{-2}$ 
and
$\nu = 0.16\ {\rm fm}^{-2}$ 
correspond to
$b =  1,67\ {\rm fm}$ and $1.77\ {\rm fm}$, respectively.
In Fig.~\ref{nu-dep}, the energy minimum point is obtained at $R \sim 2$ fm
in the case of $\nu = 0.18$~fm$^{-2}$ and the optimal $R$ value for $\nu = 0.16$~fm$^{-2}$
is a bit smaller, $R \sim 1.5$ ~fm.
It looks that cluster is hindered.
However, the size of each $\alpha$ cluster is expanded, 
and the rms radius becomes bigger in the $\nu = 0.16$~fm$^{-2}$ case, which is optimal for $^{24}$O.
The matter rms radius is 2.83~fm in the case of $R = 2$ fm and $\nu = 0.18$~fm$^{-2}$,
and it increases to 2.93 fm 
in the case of $R  = 1.5$ fm and $\nu = 0.16$~fm$^{-2}$,
even though the $R$ value is smaller.
We have previously shown in Be isotopes that when neutrons stay between the $\alpha$ clusters,
$\alpha$-$\alpha$ clustering is enhanced~\cite{PhysRevC.61.044306,PhysRevC.62.034301}.
This kind of deformation allows to reduce the kinetic energy of neutrons in the $sd$ shell orbit as indicated in the Nilsson diagram.
 In the present case of Oxygen isotopes, mean field features are getting more important than in Be isotopes.
 When neutrons are added to $^{22}$O at the center,
the expansion of each $\alpha$ clusters is energetically more favored than enhancing the clustering 
for reducing the kinetic energy of the neutrons.
 
Based on the generator coordinate method (GCM), the superposition of
different Slater determinants can be done.
The ground state of $^{24}$O is obtained at $-165.80$ MeV,
where the experimental value is $-168.97$ MeV. 
The calculated rms radius of $^{22}$O and $^{24}$O are 2.75 fm and 2.92 fm, respectively.
Experimentally, the matter radius of $^{22}$O is deduced to be 2.88 fm~\cite{PhysRevLett.117.052501},
and it  increases by about 0.2 fm at $^{24}$O.
Although the radii of $^{22}$O and $^{24}$O are slightly smaller than the experimental values,
the jump at $^{24}$O from $^{22}$O is reproduced in the present analysis.


\begin{table}
 \caption{
 The optimized $\Lambda$ and $R'$ (radius parameter for the valence neutrons) values
 of $^{24}$O 
 for each $R$ (radius parameter for the four $\alpha$ clusters).
 The size parameter $\nu$ in Eq.~\ref{spwf}  is
is 0.16 fm$^{-2}$.
 }
  \begin{tabular}{ccc} \hline \hline
 $R$  (fm)  & $\Lambda$ &  $R'$ (fm) \\ \hline
 0.5           &    0.4          &     1.5    \\
 1.0           &    0.4          &     2.0    \\
 1.5           &    0.5          &     2.0    \\
 2.0           &    0.4          &     2.5    \\
 2.5           &    0.4          &     2.5    \\
 3.0           &    0.4          &     3.0    \\
\hline
  \end{tabular}   
\label{adopted-value}
\end{table}

\section{Summary}\label{summary}

The anomalous large radius of $^{24}$O, located at the dripline of Oxygen isotopes,
was examined, utilizing
two degrees of freedom for the large radius; the relative distances among
four $\alpha$ clusters and size of each $\alpha$ cluster,
where Tohsaki interaction, which has finite range three-body terms
is employed. The nucleus $^{24}$O has the almost the same amount of the clusterization 
compared with $^{22}$O, but the expansion of each $\alpha$ cluster plays an important role.
The expansion of the $^{22}$O core part has been known to be important for the large radius of $^{24}$O,
and this is naturally explained as a result of variational calculation.
The calculated rms matter radius of $^{22}$O and $^{24}$O are 2.75 fm and 2.92 fm, respectively.
Although these are slightly smaller than the experimental values,
the jump at $^{24}$O from $^{22}$O is reproduced.
We have previously shown that when neutrons stay between the $\alpha$ clusters,
$\alpha$-$\alpha$ clustering is enhances in Be isotopes.
This kind of deformation allows to reduce the kinetic energy of neutrons in the $sd$ shell orbit as indicated in the Nilsson diagram.
 In the present case of Oxygen isotopes, 
 mean field features are getting more important than in Be isotopes.
When neutrons are added to $^{22}$O at the center,
the expansion of each $\alpha$ clusters is energetically more favored than enhancing the clustering 
for reducing the kinetic energy of the neutrons.

\begin{acknowledgments}
Numerical calculation has been performed using the computer facility of 
Yukawa Institute for Theoretical Physics,
Kyoto University. This work was supported by JSPS KAKENHI Grant Number 17K05440.
\end{acknowledgments}

\bibliographystyle{apsrev4-1}
\bibliography{biblio_ni}

%
%

\end{document}